# Alternative approach to the description of metallic clusters

Yuri Kornyushin

*Maître Jean Brunschvig Research Unit, Chalet Shalva, Randogne, CH-3975*

A detailed simple model is applied to study a metallic cluster. It is assumed that the ions and delocalized electrons are distributed randomly throughout the cluster. The delocalized electrons are assumed to be degenerate. A spherical ball models the shape of a cluster. The energy of the microscopic electrostatic field around the ions is taken into account and calculated. It is shown in the framework of the model that the cluster is stable. Equilibrium radius of a ball and the energy of the equilibrium cluster are calculated. Bulk modulus of a cluster is calculated also.

## 1. Introduction

A double-jelly model (delocalized electrons jelly and ions jelly) was applied for the description of metallic clusters [1]. This model does not take into account microscopic electrostatic field around the ions. It was shown recently that this field contributes essentially to the energetic balance of a system [2]. Here we shall take into account this microscopic field. In [2] it was considered a case when the delocalized electrons are classical. Here we shall analyze the degenerate delocalized electrons.

A spherical ball models the shape of a cluster. We shall model the conditions as adiabatic ones under given pressure. We shall restrict our consideration here by given entropy and given pressure condition. In this case the thermodynamic potential, having a minimum in the state of equilibrium, is the enthalpy (it is a function of thermodynamic variables, entropy $S$ and pressure $P$: $H(S,P)$ [3]. Anyway under different condition the contribution of the entropy term to the Gibbs free energy is negligible.

We shall consider here only atmospheric pressure, which can be neglected. So really, we will take into account the energy of a cluster only.

## 2. Electrostatic energy of a separate ion

Let us consider a ball of a metallic cluster of a volume $V = 4\pi R^3/3$ ($R$ is the radius of a cluster), consisting of $n$ ions and $zn$ delocalized electrons. We consider here the ions as point charges, and the delocalized electrons like a negatively charged gas. In metallic clusters the delocalized electros density is so high, that they are degenerate over all the temperature range of the existence of metallic cluster. Degenerate delocalized electrons screen a long-range electrostatic field of point charges. The screening Thomas-Fermi radius is as follows [4]:

$$1/g = (VE_F/6\pi zne^2)^{1/2} = 0.4714 R^{3/2}(E_F/zn)^{1/2}/e = R^{3/2}/R_0^{1/2}, \qquad (1)$$

where $e > 0$ is the elementary charge, $R_0 = 4.5 zne^2/E_F$, and $E_F = (3\pi^2 zn/V)^{2/3}(\hbar^2/2m)$ is the Fermi energy [4]. Here $m$ is the electron mass and $\hbar$ is Planck's constant, divided by $2\pi$.

The electrostatic field around a separate positive ion, submerged into the gas of degenerate delocalized electrons, is as follows [4]:

$$\varphi = (ze/r)\exp{-gr}, \qquad (2)$$

where $r$ is the distance from the center of the ion.

The electrostatic energy of this field is the integral over the ball volume of its gradient in a second power, divided by $8\pi$. The lower limit of the integral on $r$ should be taken as $r_0$, a very small value. Otherwise the integral diverges. Calculation yields the following expression for the electrostatic energy of a separate ion:

$$U_0 = 0.5 z^2 e^2 (r_0^{-1} + 0.5g) \exp{-2gr_0}. \qquad (3)$$

For $2gr_0$ essentially smaller than unity Eq. (3) yields:

$$U_0 = (z^2 e^2 / 2r_0) - 0.75(z^2 e^2 g), \; g = R_0^{1/2}/R^{3/2}. \qquad (4)$$

The first term in the right-hand part of Eq. (4), $z^2 e^2 / 2r_0$, represents the electrostatic energy of the bare ion.

It is worthwhile to note that the expansion of a cluster leads to the decrease in the delocalized electron density. From this follows the increase in the screening radius [see Eq. (1)]. The electrostatic energy of a separate ion increases concomitantly. One can see it, regarding Eq. (4).

### 3. Total energy of a cluster

We regard the ions of the cluster as randomly distributed. It is well known since 1967, that the electrostatic energy of $n$ randomly distributed ions is just $U = nU_0$ [5].

The kinetic energy, $T$, of a gas of delocalized electrons, confined in a volume $(4/3)\pi R^3$, and calculated in accord with the ideas of the Thomas-Fermi model, is as follows [1]:

$$T = 1.105 (zn)^{5/3} (\hbar^2 / mR^2), \qquad (5)$$

So the total energy of a cluster, $W(R)$, is as follows:

$$W(R) = 1.105 (zn)^{5/3} (\hbar^2 / mR^2) + (z^2 e^2 n / 2r_0) - 0.75(z^2 e^2 n R_0^{1/2}/R^{3/2}), \qquad (6)$$

As a function of $R$, $W(R)$ has a minimum. So, the cluster is stable. He minimum value of $W(R)$ is

$$W_e = (n z^2 e^2 / 2r_0) - 0.565 z^{7/3} n (me^4/\hbar^2). \qquad (7)$$

This minimum occurs when $R = R_e$:

$$R_e = 2.422 (n/z)^{1/3} (\hbar^2 / me^2). \qquad (8)$$

The equilibrium volume per one ion is

$$v_e = 4\pi R_e^3 / 3n = 59.532 (\hbar^2/me^2)^3 / z. \qquad (9)$$

For $z = 1$ $v_e = 8.821 \times 10^{-24}$ cm$^3$. It is about 3 times smaller than expected.

### 3. Bulk modulus of a cluster

When condensed matter subject is expanded, the increase in its elastic energy is as follows [6]:

$$\delta W = 0.5K(\delta V)^2/V_0 = 4.5KV_0(\delta R/R_0)^2, \tag{10}$$

where $K$ is the bulk modulus, $V_0$ is the initial volume, and $R_0$ is the initial radius. Here a well-known relation, $\delta V/V_0 = 3\delta R/R_0$, was used. It is valid when $\delta R$ is essentially smaller than $R_0$.

The change in the total energy of a cluster [Eq. (6)] is

$$\delta W = 0.5(\partial^2 W/\partial R^2)_{R = Re}(\delta R)^2. \tag{11}$$

Taking into account that $R_0$ and $R_e$ are the values of the same meaning, it follows from Eqs. (10,11), that

$$K = (1/12\pi R_e)(\partial^2 W/\partial R^2)_{R = Re}. \tag{12}$$

Eqs. (6,8,12) yield the following expression for the bulk modulus:

$$K = 0.00105(z^{10/3}m^4 e^{10}/\hbar^8). \tag{13}$$

For $z = 1$ Eq. (13) yields $K = 3.102 \times 10^{11}$ erg/cm$^3$. This value is a very reasonable one indeed.

## 4. Discussion

We modeled the shape of metallic cluster here as a spherical ball. The delocalized electrons in metallic cluster are degenerate. Their kinetic energy is calculated in the spirit of the Thomas-Fermi model. The delocalized electrons screen the electrostatic field of the ions. This field was calculated; it was shown that it contributes essentially to the energy of the cluster.

The equilibrium values of the energy of the cluster and its volume were calculated. The bulk modulus of a cluster was calculated also. Its value appears to be quite reasonable.